# High-temperature thermoelectric properties of novel layered bismuth-sulfide $LaO_{1-x}F_xBiS_2$

Atsushi Omachi, Joe Kajitani, Takafumi Hiroi, Osuke Miura, Yoshikazu Mizuguchi*

Department of Electrical and Electronic Engineering, Tokyo Metropolitan University, 1-1, Minami-osawa, Hachioji, 192-0397, Japan

Corresponding author: Yoshikazu Mizuguchi (mizugu@tmu.ac.jp)

Abstract

We have investigated the high-temperature thermoelectric properties of the layered compound $LaO_{1-x}F_xBiS_2$. The electrical resistivity of $LaOBiS_2$ showed an anomalous behavior; a metal-semiconductor transition was observed around 270 K. It was found that the value of the electrical resistivity decreased with F substitution. The Seebeck coefficient decreased with increasing F concentration. The highest power factor of 1.9 $\mu W/cmK^2$ at 480 °C was obtained for $LaOBiS_2$.

## 1. Introduction

Discovery of novel thermoelectric material is one of the most important issues for development of thermoelectric application [1]. Particularly, layered materials have been actively studied because their low-dimensional electronic states could result in high thermoelectric properties as realized in $Bi_2Te_3$, cobalt oxides and $CsBi_4Te_6$ [2-4]. Recently, layered compounds with $BiS_2$ layers have got much attention because two-dimensional superconductivity with transition temperature as high as 10.6 K was observed [5-13]. Among the $BiS_2$-based layered compounds, $LaOBiS_2$ is the typical system, and it has mostly studied in this one year [6,11]. The schematic image of the crystal structure of $LaOBiS_2$ is shown in Fig. 1. It is composed of an alternate stacking of the conduction layer ($BiS_2$ layer) and the blocking layer (LaO layer). Recently, we found that the electrical resistivity shows an anomalous behavior. Figure 2 shows the temperature dependence of resistivity for the polycrystalline sample of $LaOBiS_2$. The resistivity decreases on cooling above 270 K while the resistivity shows a semiconducting behavior below 270 K. Because of the layered structure and the observed anomalous characteristics on the electrical conductivity, we have expected that pristine or carrier-doped $LaOBiS_2$ could be a potential candidate for novel thermoelectric materials. Here, we show the thermoelectric properties at high temperatures for $LaO_{1-x}F_xBiS_2$.

## 2. Methods

Polycrystalline samples of $LaO_{1-x}F_xBiS_2$ were prepared by the solid-state reaction method using powders of $La_2S_3$ (99.9 %), $Bi_2O_3$ (99.9 %), $BiF_3$ (99.9 %), $Bi_2S_3$ and grains of Bi (99.99 %). The $Bi_2S_3$ powder was synthesized by reacting Bi (99.99 %) and S (99.99 %) grains at 800 ºC in an evacuated quartz tube. Other chemicals were purchased from Kojundo-Kagaku Laboratory. The mixture of starting materials with nominal compositions of $LaO_{1-x}F_xBiS_2$ ($x$ = 0, 0.05, 0.25, 0.5) was well mixed, pelletized and sealed into an evacuated quartz tube. The $LaO_{1-x}F_xBiS_2$ pellets were heated at 800 ºC for 15h. The obtained samples were ground, pelletized, sealed into an evacuated quarts tube and heated under the same heating conditions to homogenize the samples.

The prepared samples were characterized by powder x-ray diffraction with CuKα radiation using RIGAKU Smart-Lab. The electrical resistivity and the Seebeck coefficient were measured by the four-terminal method using ULVAC-RIKO ZEM-3 up to 480 ºC in an atmosphere of low-pressure He gas.

*3. Results and discussion*

Figure 3 shows the powder x-ray diffraction patterns for LaO$_{1-x}$F$_x$BiS$_2$. Almost all of the peaks were indexed using the tetragonal space group of *P*4/*nmm*. For $x$ = 0 and 0.05, small peaks of the impurity La$_2$O$_2$S were detected as indicated with the "+" symbols in Fig. 3. It should be noted that the La$_2$O$_2$S impurity is insulator. Hence, the impurity phase does not affect the thermoelectric properties. The lattice constants *a* and *c* were calculated using the peak positions of the (200) and (004) peaks. The calculated lattice constants are displayed in Fig. 4. With increasing F concentration, the length of the *c* axis continuously decreases while the length of the *a* axis does not show a remarkable change. The changes in the lattice constants upon F substitutions are consistent with the previous experiments [14,15].

Figure 5 shows the temperature dependence of the electrical resistivity for LaO$_{1-x}$F$_x$BiS$_2$. With increasing F concentration, the value of resistivity decreases at whole temperatures. This fact is consistent with the scenario that the electron carriers could be generated in the BiS$_2$ layers upon F substitution. For $x$ = 0, the resistivity increases with increasing temperature, which is a metallic behavior. For $x$ = 0.05 – 0.5, the temperature dependence of the resistivity does not show small remarkable changes with increasing temperature. For all the samples, humps are observed as indicated with the triangle symbols in Fig. 5. The hump shifts to a higher temperature with increasing F concentration. The hump may indicate the evolution of ordered states something like charge-density-wave state which have been theoretically predicted to be possible to occur in the BiS$_2$-based family.

Figure 6 shows the temperature dependence of the Seebeck coefficient for LaO$_{1-x}$F$_x$BiS$_2$. For all the samples, the Seebeck coefficient increases with increasing temperature. The value of the Seebeck coefficient decreases with increasing F concentration. The highest Seebeck coefficient of 200 μV/K was observed at 480 °C for $x$ = 0.

The power factor was calculated using the equation of $P = S^2/\rho$ (*P*: power factor, *S*: Seebeck coefficient, $\rho$: electrical resistivity). Figure 7 shows the temperature dependence of the power factor for LaO$_{1-x}$F$_x$BiS$_2$. For all samples, the power factor increases with increasing temperature. With increasing F concentration, the power factor decreases. The highest power factor of 1.9 μW/cmK$^2$ at 480 °C is obtained for $x$ = 0.

Although the obtained power factor of LaOBiS$_2$ is much lower than that of thermoelectric materials used in a practical application and the partial F substitution is not effective for enhancing power factor, we believe that further improvement of the power factor in this material is possible. Since this material possesses a layered structure, we will be able to tune the physical properties by manipulating the blocking layer. A partial or full substitution of the La site by other lanthanide elements, such as La$_{1-x}$$Ln$$_x$OBiS$_2$ ($Ln$ = lanthanide), will generate a chemical pressure effect. Further, a partial substitution of La by alkaline earth elements, such as La$_{1-x}$$A$$_x$OBiS$_2$ ($A$ = alkaline earth metal) will provide hole carriers in the BiS$_2$ layers. Another strategy is a replacement of the blocking layer structure. If a blocking layer thicker along the $c$ axis could be inserted between conduction layers, two-dimensionality should be enhanced and the thermoelectric properties may be enhanced.

4. Conclusion

We have synthesized the polycrystalline samples of new layered bismuth-sulfide LaO$_{1-x}$F$_x$BiS$_2$. To investigate the thermoelectric properties, the electrical resistivity and Seebeck coefficient at high temperature up to 480 ºC were measured. LaOBiS$_2$ showed a metallic behavior of electrical resistivity above 270 K while it showed a semiconducting behavior below 270 K. It was found that the value of the electrical resistivity decreased with F substitution. The Seebeck coefficient decreased with increasing F concentration. The highest power factor of 1.9 μW/cmK$^2$ at 480 ºC was obtained for LaOBiS$_2$. Although the obtained power factor was lower than those of the thermoelectric materials on practical use, we will be able to improve the thermoelectric properties by manipulating the blocking layer in the BiS$_2$-based layered family.

*Acknowledgements*

The authors thank Dr. H. Takatsu for experimental supports and fruitful discussion. This work was partly supported by a Grant-in-Aid for Scientific Research for Young Scientist (A) and a research fund from The Thermal and Electric Energy Technology Foundation.

Figure captions

Fig. 1. Schematic image of the crystal structure of LaOBiS$_2$.

Fig. 2. Temperature dependence of the resistivity for LaOBiS$_2$.

Fig. 3. Powder x-ray diffraction patterns for LaO$_{1-x}$F$_x$BiS$_2$. The numbers shown with $x =$ 0.5 are Miller indices. The peaks of impurity La$_2$O$_2$S are indicated with "+".

Fig. 4. (a) F concentration dependence of lattice constant $a$ for LaO$_{1-x}$F$_x$BiS$_2$. (b) F concentration dependence of lattice constant $c$ for LaO$_{1-x}$F$_x$BiS$_2$.

Fig. 5. Temperature dependence of electrical resistivity for LaO$_{1-x}$F$_x$BiS$_2$. The triangle symbol indicates the temperature of the apical of the anomaly.

Fig. 6. Temperature dependence of the Seebeck coefficient for LaO$_{1-x}$F$_x$BiS$_2$.

Fig. 7. Temperature dependence of the Power factor for LaO$_{1-x}$F$_x$BiS$_2$.

Fig. 1.

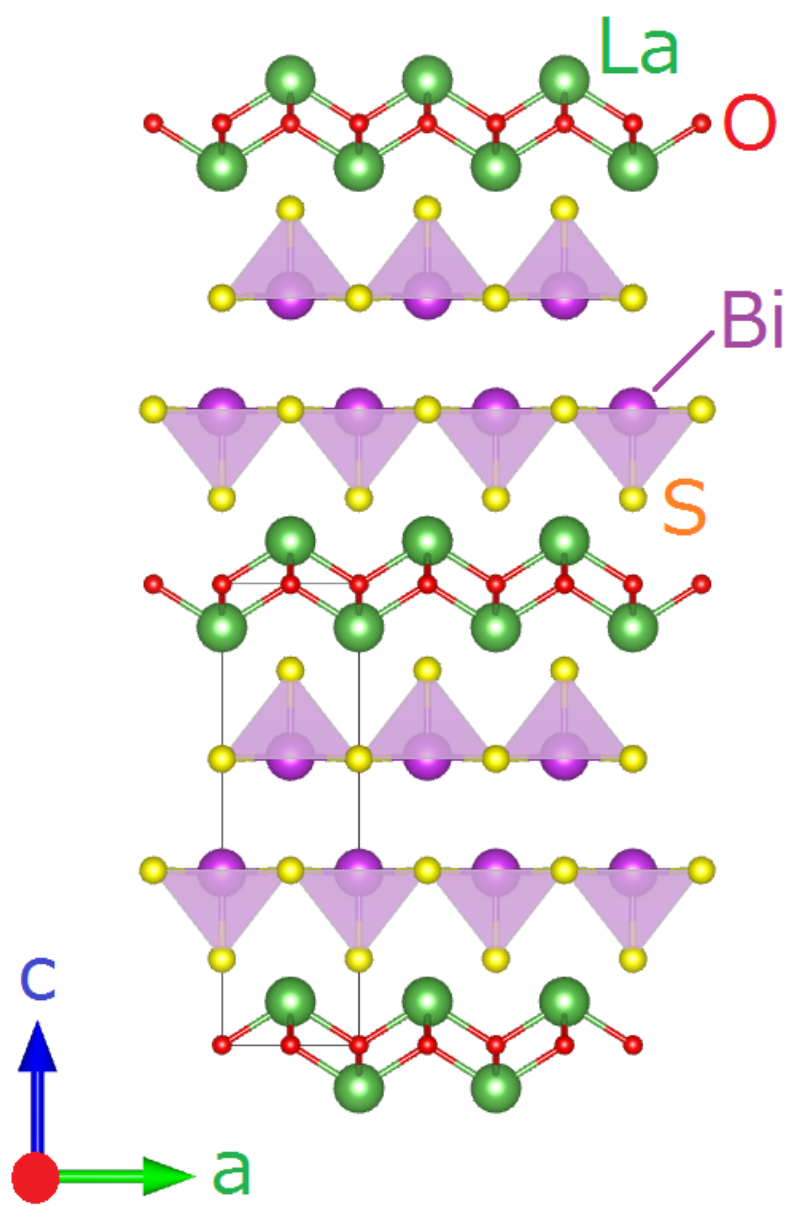

Fig. 2.

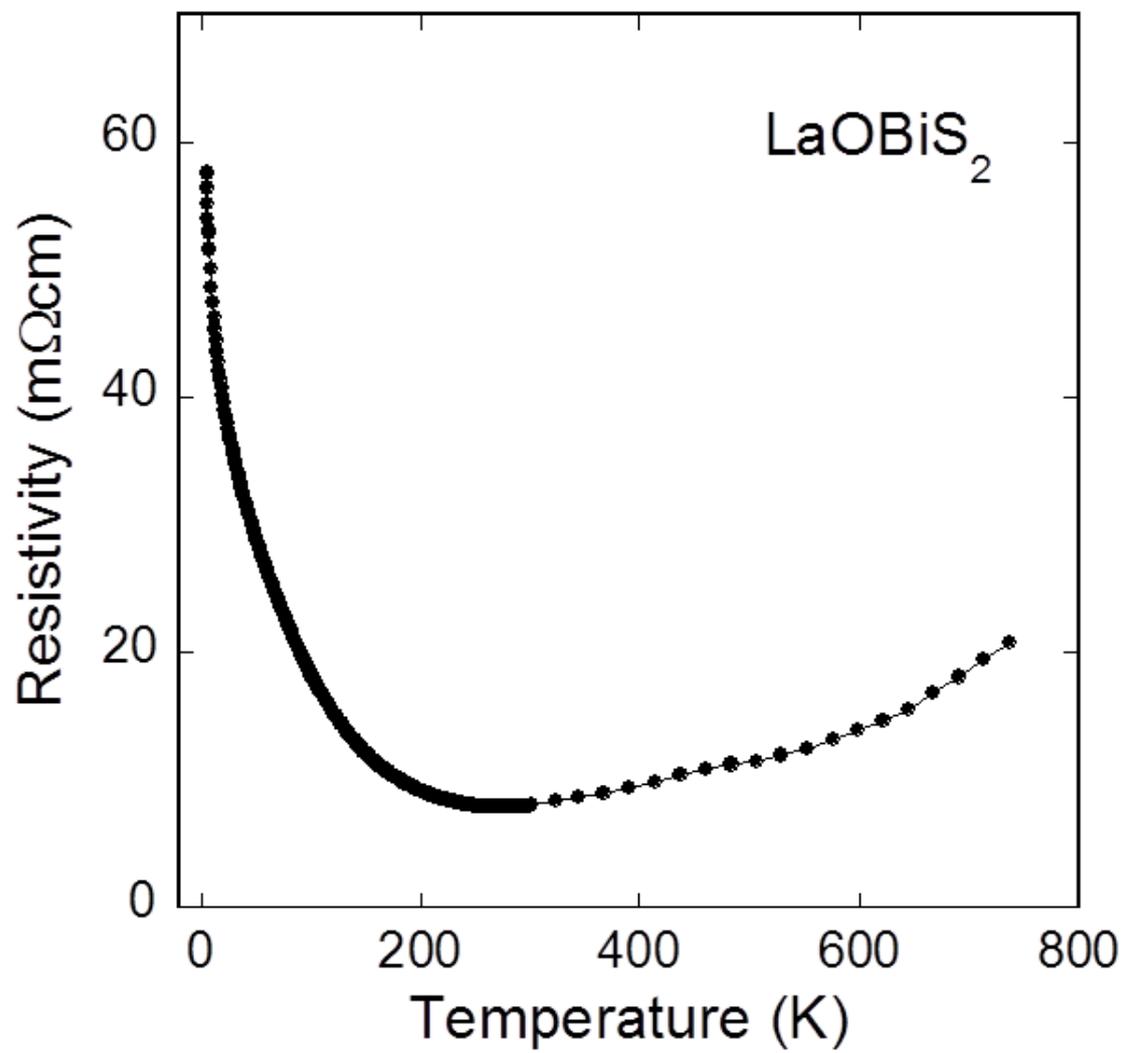

Fig. 3.

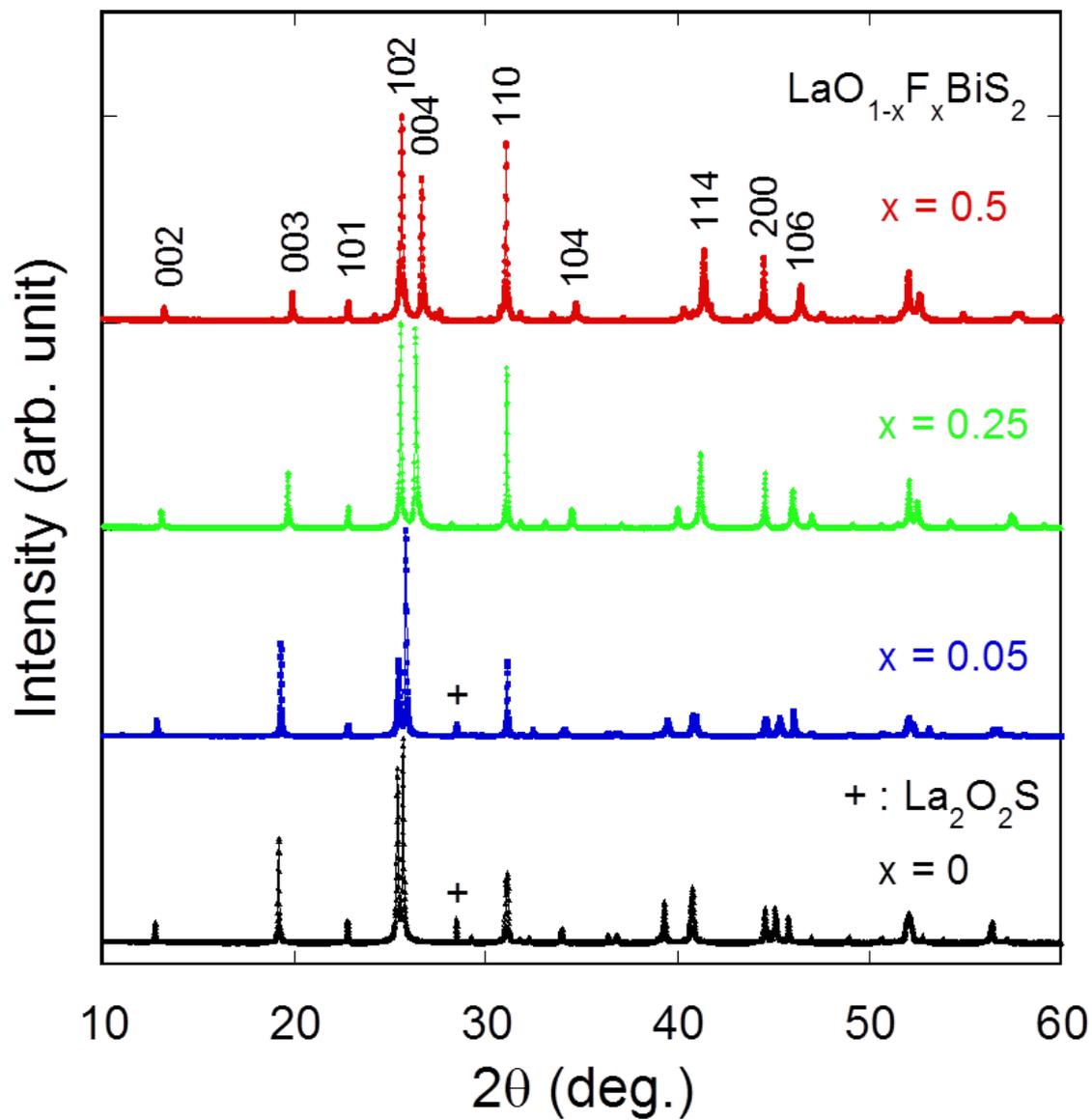

Fig. 4.

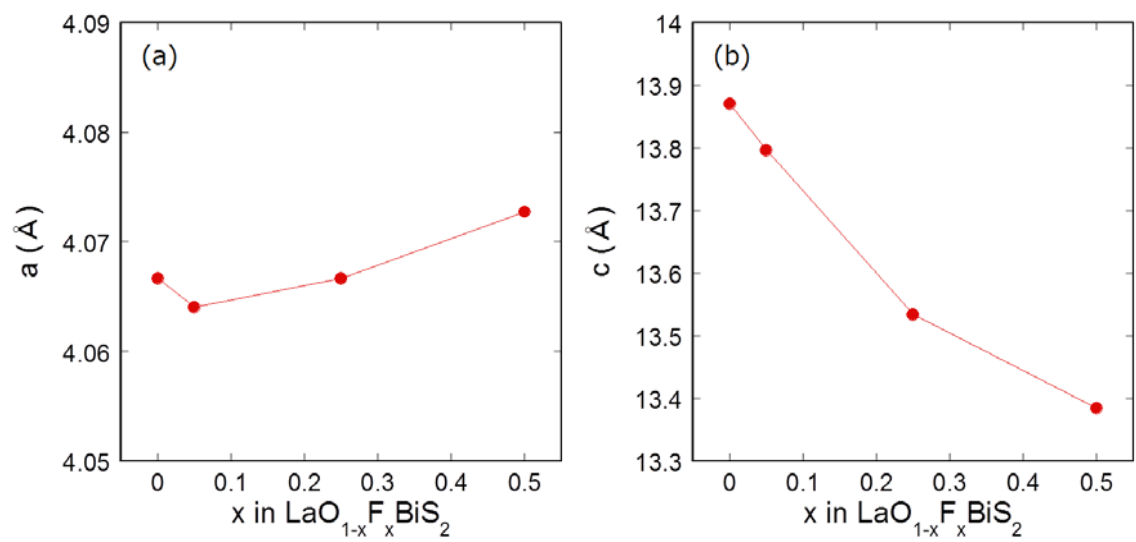

Fig. 5.

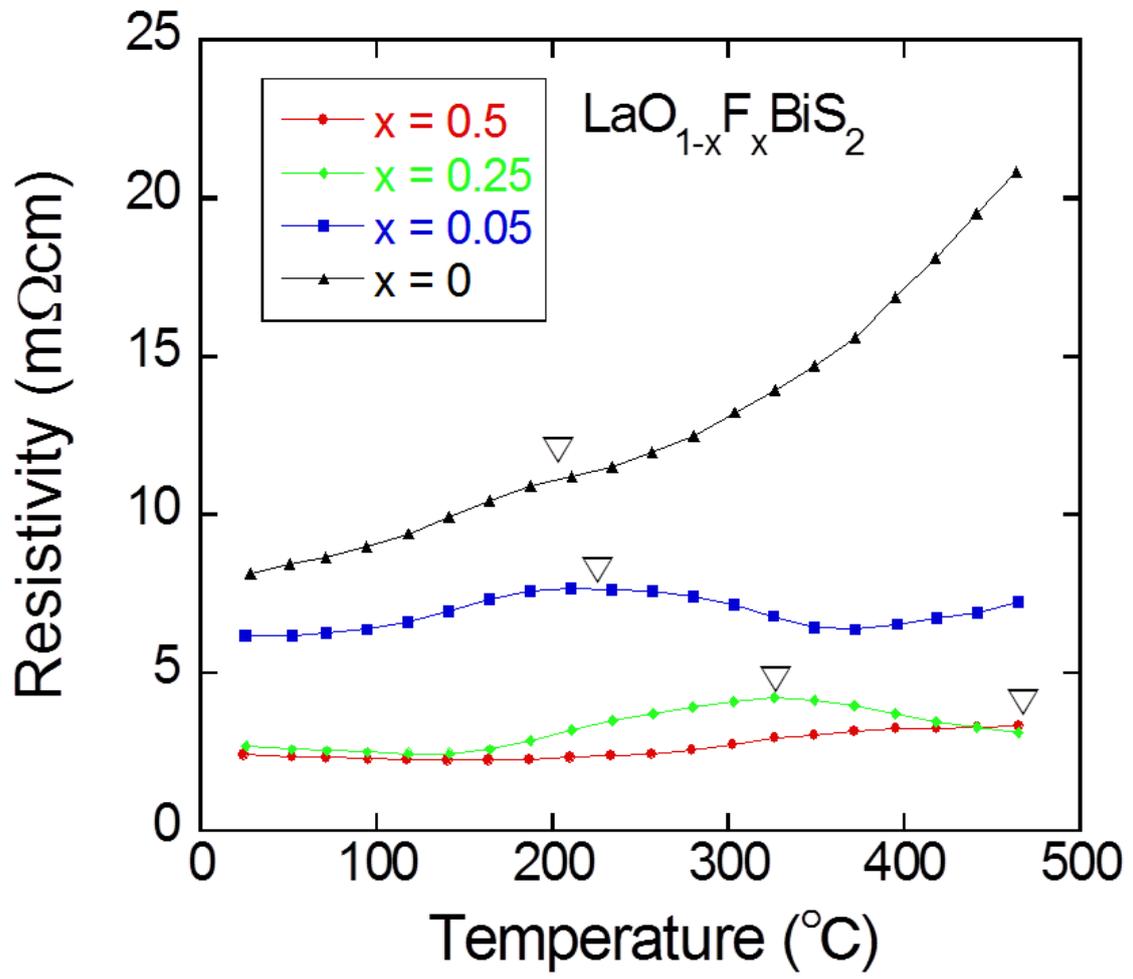

Fig. 6.

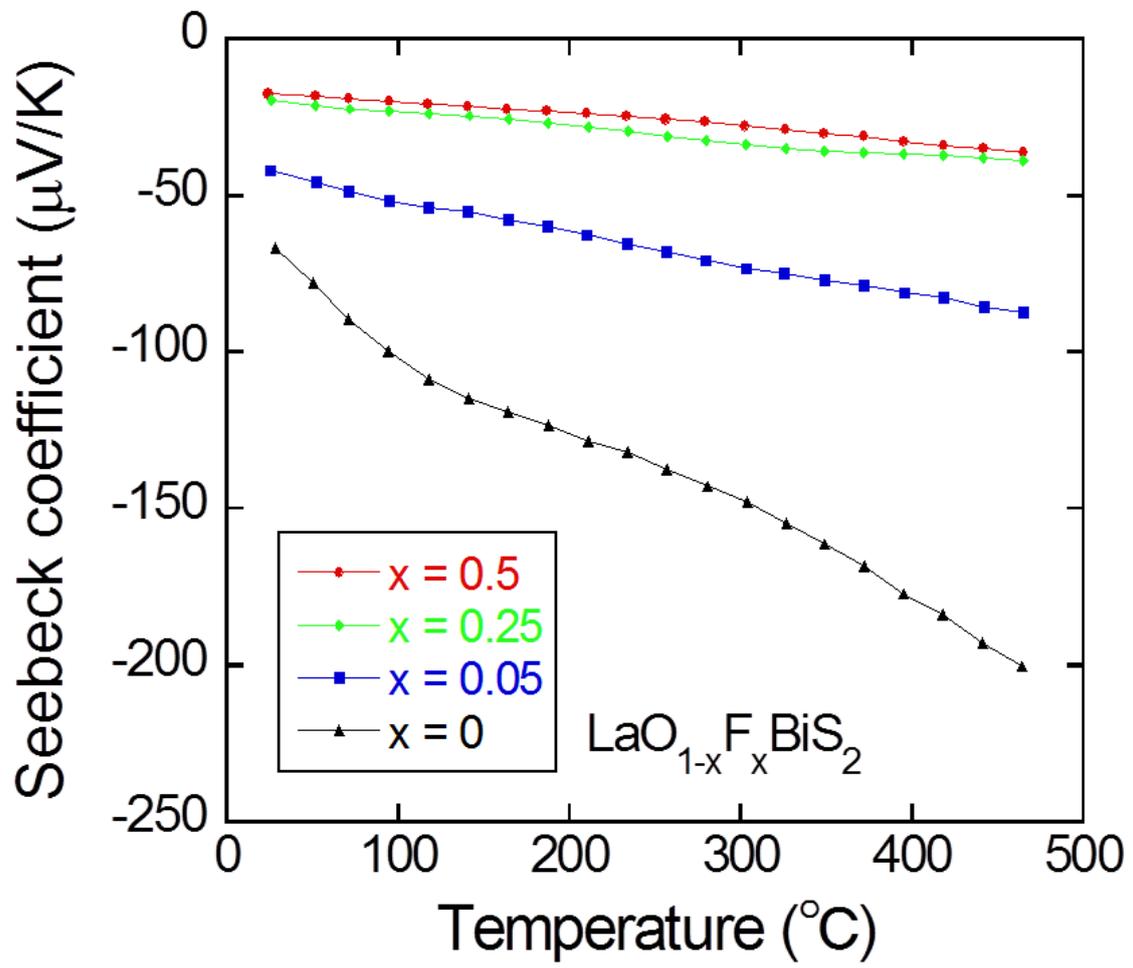

Fig. 7.

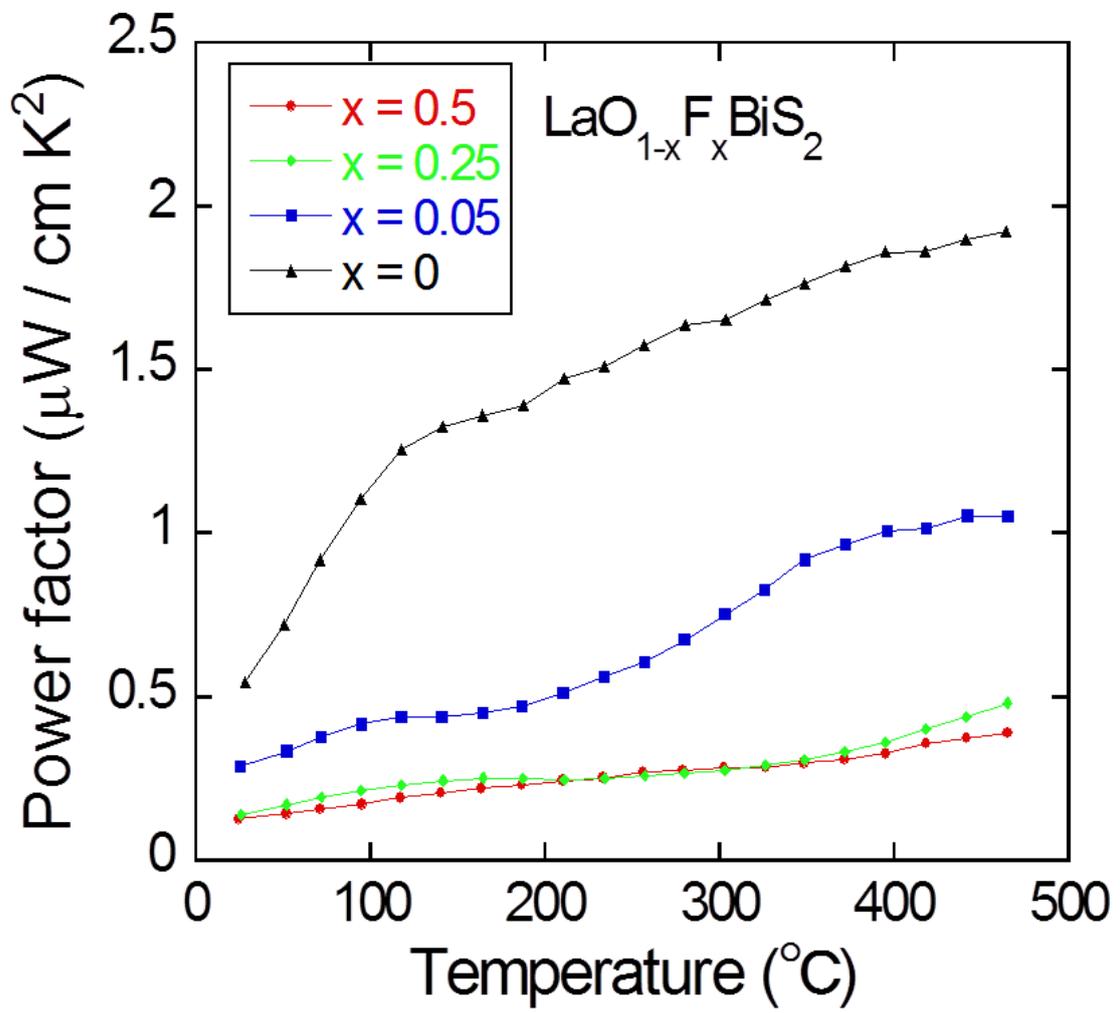